# MAPPING EXCELLENCE IN NATIONAL RESEARCH SYSTEMS: THE CASE OF ITALY[1]

GIOVANNI ABRAMO

*Laboratory for Studies of Research and Technology Transfer, Dept of Management, School of Engineering, University of Rome "Tor Vergata" – Italy*
*and*
*Italian Research Council*

CIRIACO ANDREA D'ANGELO, FLAVIA DI COSTA

*Laboratory for Studies of Research and Technology Transfer, Dept of Management, School of Engineering, University of Rome "Tor Vergata" - Italy*

*The study of the concept of "scientific excellence" and the methods for its measurement and evaluation is taking on increasing importance in the development of research policies in many nations. However, scientific excellence results as difficult to define in both conceptual and operational terms, because of its multi-dimensional and highly complex character. The literature on the theme is limited to few studies of an almost pioneering character. This work intends to contribute to the state of the art by exploring a bibliometric methodology which is effective, simple and inexpensive, and which further identifies "excellent" centers of research by beginning from excellence of the individual researchers affiliated with such centers. The study concentrates on the specific case of public research organizations in Italy, analyzing 109 scientific categories of research in the so called "hard" sciences and identifying 157 centers of excellence operating in 60 of these categories. The findings from this first application of the methodology should be considered exploratory and indicative. With a longer period of observation and the addition of further measurements, making the methodology more robust, it can be extended and adapted to a variety of national and supranational contexts, aiding with policy decisions at various levels.*

The theme of scientific excellence in research is of ever-increasing importance in the field of research policy, stimulating interest even among those that are not directly implicated in the field. The ability to successfully identify national centers of excellence would clearly allow higher allocative efficiency in research funding. At the same time it would reduce "information asymmetries" between the public and private sectors in selecting and forming partnerships, and generally aid in supply-demand relationships between parties. At supra-national levels, such as that of the European Union, further motivations for identifying excellence are: to rationalize R&D spending, avoiding scattered and redundant efforts; and to favor integration and synergies leading to a critical mass of research that can compete at the global level.

---

[1] Abramo, G., D'Angelo, C.A., Di Costa, F. (2009). Mapping excellence in national research systems: the case of Italy. *Evaluation Review*, 33(2), 159-188. DOI: 10.1177/0193841X08322871

But the concept of scientific excellence, seemingly intuitive, actually results as multi-dimensional and highly complex in character. It is difficult to agree on a conceptual definition and even more difficult to apply the concept in operation, due particularly to a failure to agree on methods of measurement. The European Commission document "How to map excellence in research and technological development in Europe" (EC 2001) has indicated some general guidelines as to the requisites of a methodology for correctly identifying centers of excellence:

- *reliability and robustness,* in relation to the differing categories of science and for the various European nations;
- *objectivity,* with mapping based as much as possible on irrefutable quantitative data that offer comparability among organizations of diverse nature, situated in diverse national systems;
- *transparency,* concerning input data, methods used and results obtained;
- *suitability for repeated use over time,* with the objective of furnishing a dynamic portrait of the evolution of excellence over time.

The simultaneous satisfaction of these requirements results as highly complex, especially in a methodology that also permits broad census exercises within reasonable times and costs. Thus the studies reported in literature have so far almost always concerned methodologies and investigations of limited focus, analyzing only one or a few scientific categories. What is more, these studies are at the "*meso*" level, meaning investigation carried out at the level of entire research organizations (EC-DGR 2004; CWTS-Fraunhofer 2003; Tijssen 2003; Van Leeuwen, Visser, Moed, Nederhof, and Van Raan 2003; Tijssen, Visser, and Van Leeuwen 2002; NEG 2001). The present work adds to the investigations into approaches for mapping excellence by exploring an innovative methodology of identifying centers of excellence. The methodology is of a "bottom up" type, meaning that it is based on the analysis of performance of single scientists: it includes a census of the entire production of scientists in Italy over a three year period, in 6 of the 8 scientific macro-areas indexed by Thomson Scientific SCI™[1] (see appendix); covering all the researchers on staff at all public research organizations. With respect to the state of the art, the current work is notable for:

i) the broad field of observation, and its exhaustive nature, consisting of all organizations in the Italian public research system (universities, public research laboratories, research hospitals);
ii) the amplitude of the set of scientific disciplines considered (109 categories in 6 macro-areas indexed by the SCI™). For the complete list see appendix;
iii) the detail of analysis of excellence (to the level of single scientists).

This type of analysis can be considered a support to a variety of figures in national and international research systems:

i) for the European Community it provides a direct contribution to mapping excellence, by discipline and by geographic distribution, and thus to consolidating and strengthening the "European research space";
ii) for national and regional policy makers, the proposed method can support allocative efficiency in pubic funding for research organizations;
iii) for single private enterprises, mapping of excellence reduces asymmetric information in the market of university industry research collaborations, rendering the choice of partnerships easier and more efficient;
iv) for public research organizations, mapping of excellence provides a system of benchmarking, which can stimulate improvements in research efficiency;

v) at the level of single individuals and organizations, the analysis can confer international prestige and visibility, as well as aiding in the mobility and networking of top scientists among centers of excellence.

Following this introductory section, the next section of the article continues with an analysis of the literature on the argument. A third section provides a detailed presentation of the methodology adopted, while the fourth section illustrates the main findings from its application. The last section presents the authors' conclusions.

## MAPPING OF CENTERS OF EXCELLENCE: MAJOR RECURRING ISSUES

The problem of identifying "scientific excellence" is a challenging task for those scholars active in the measurement and evaluation of scientific research. In recent years, international literature has featured an abundance of methodological contributions concerning the general subject of evaluation (such as Adams, and Griliches 1996; Moed Glänzel, and Schmoch 2004; Van Raan 2005; Abramo, and D'Angelo 2007). However, the specific problem of mapping out centers of excellence among research organizations in national or supranational systems has distinct features and requires unique methodologies when compared to other forms of evaluation, such as the evaluation of the simple efficiency of the scientific activity of organizations.

Tijssen, Visser, and Van Leeuwen (2002), in assessing research performance of Dutch universities active in the principal scientific disciplines, propose a solution using an indicator termed the "Highly Cited Papers Index", or HCP index. Among the university research groups, they identify centers of excellence as being those which achieve high concentrations of publications with top HCP. Also, their analysis further illustrates that researchers affiliated with centers of excellence tend to submit their papers to the most prestigious international journals.

To capture the entire range of multiple characteristics of excellence, Tijssen (2003) suggests a systematic and interactive approach that uses quantitative indicators, including bibliometric indicators, and a wide range of further information in a "scoreboard" framework. The publication reports a study in the identification of centers of excellence in faculties of economics at Dutch universities.

But the necessity of identifying research of the highest quality entails a step up in technique: from average value measures of productivity (based on technical efficiency scores) and impact (based on bibliometric impact scores), to techniques using indicators that single out concentrations of results in the upper extreme of performance indicator distribution curves. Techniques must point out centers that achieve concentrations of highly cited or "top" articles. Towards this objective, Van Leeuwen, Visser, Moed, Nederhof, and Van Raan (2003), propose the combined use of various bibliometric indicators that each reflect a dimension of the scientific impact of publications from individual research organizations. The study provides an example by carrying out an analysis of performance of Dutch universities that are active in chemistry research .

A series of studies towards mapping centers of excellence have also been carried out at the European Community level, to assist with policy and structuring for the "European Research Space". Three published studies concentrate on the areas of life sciences (CWTS-Fraunhofer 2003), nanotechnology (NEG 2001) and economics (EC-DGR 2004). The first of these studies is entirely based on analysis of output by identified organizations in terms of publications and patents. The second reports an analysis of output, but integrated with a survey of the reputation of the organizations. The third conducts both bibliometric analyses and reputation assessments, but also measures structural dimensions such as funding and human resources. As a final indicator of excellence it also measures the number of projects that centers have activated within the Marie Curie Fellowships Program or the EU Fifth European Framework Program.

The overall literature analysis reveals that there is a clear trend among scholars to adopt multi-factorial models of measurement of a highly articulated character. There is a convergence towards considering excellence as a multi-dimensional construct, for which measurement must examine diverse variables. These variables deal above all with output (scientific publications, patents, etc.) and seem to concentrate on several proxies for measuring their "value", especially bibliometric indicators of varying sophistication (while not excluding other indicators, such as judgments based on expert opinion).

A concern that should be noted is that the contributions found in the literature almost always report an analysis that focuses on only one or a very few scientific categories, often providing analysis of only one or a very few scientific categories. They also always report on a *meso* level of investigation, meaning an examination of output of entire research organizations. There are no studies based on levels of analysis of greater detail, such as of single laboratories, sections, or research groups within the organizations themselves. Yet it is clear that measuring from an aggregate institutional basis can hide specific cases of excellence: smaller research groups of devoted and coordinated scientists that achieve goals and results far superior to the average of their colleagues operating in the same organization.

With the objective of filling these gaps, the authors of the current work have developed a methodology for mapping excellence that is based on a "bottom-up" approach. This refers to an approach that starts by identifying excellence at the most fundamental level possible: the level traceable to single scientists. The approach is also sufficiently practical to permit analysis of a full range of hard-science disciplines.

## METHODOLOGICAL APPROACH AND DATASET

The first problem for the authors was developing a working definition for identifying centers of scientific research excellence.

For nations with highly articulated and large technological-scientific infrastructures (such as Italy, which is eighth among OECD nations for overall spending on R&D) there is a significant and clear trade-off in identifying centers of excellence either by multi-dimensional measurement accomplished through sampling, or by methods which permit coverage of the total population. The authors decided in favour of the second approach, thus foregoing the approach of measuring a set of multiple qualitative-quantitative indicators (potentially including level of internationalization, intensity of collaboration with the private sector, capacity to attract external financing, acknowledgement of prestige, etc.). The multi-dimensional approach would have required data collection by means of interviews with single research units (where these actually measure and record such data). A full census approach, rather than sampling analysis, permitted operating in a non-invasive manner from a central location, and also permitted avoiding well-known problems concerning interpretation of questions, homogeneity of responses and reliability of inferential analyses.

A second methodological problem concerned defining a minimum size for the organizational units to be identified as a center of excellence. This is a choice which must include consideration of the institutional context of the study being undertaken. In Italy, public research is conducted in universities, public research laboratories and research hospitals. Universities typically feature a wide spectrum of scientific disciplines but with research units of small dimensions, which are suited to educational needs. In contrast, public research laboratories are generally more specialized and so conduct research in a more limited number of scientific categories, but with much larger operative units. Defining a research group as "excellent" includes a requirement that it have a minimum number of scientists with productivity and impact at the top of the performance distribution in its given category. But in a given scientific category, all else being equal, large organizations have higher probability than small organizations of employing higher numbers of top-performers. The authors determined that they would set the presence of 4 top performers as the minimum threshold for defining a center of excellence, since in their judgment (see below) this represents a suitable compromise between meeting the demands of both significance and representativeness, while avoiding the penalization of small internal research units found in universities.

The third problem concerned the definition of a highly significant indicator of excellence, which would indeed permit a census type investigation, at a distance, with low costs and quick times for completion. In light of a trade-off between significance of the indicator and costs of obtaining data, the choice was made to opt for a bibliometric type of indicator. Judging from literature on the theme, this choice results as being the best for satisfying the two objectives of limiting complexity and achieving survey significance: the argument is that bibliometric indicators provide the most reliable proxy for measurement of scientific excellence, while other types of single variables (such as intensity of collaboration with industry, or capacity to attract external financing) appear less significantly correlated to scientific excellence (Abramo, D'Angelo, and Di Costa 2008; Abramo, D'Angelo, Di Costa, and Solazzi 2008). The authors extracted articles and reviews, published by scientists affiliated with Italian public research organizations, from the Thomson Scientific "Science Citation Index" (SCI™, CD-Rom version) for the years 2001-2003[2].

The Thomson Scientific "Journal Citation Reports"® further give the "impact factor" of each publishing journal[3]. In research assessment and academic evaluation, this impact factor is often used to provide a gross approximation of the prestige of journals in which scientists succeed in placing their publications. It is essential to make informed, careful use of this impact factor data: they should not be used without careful attention to a number of phenomena that can influence citation rates. A notable phenomenon is that the average number of citations to other scientific publications can vary in the articles originating from different fields of research. The frequency of citations in the individual articles of a single journal can also vary considerably, meaning that impact factor is generally the average value of a much skewed distribution. The use of impact factor as a proxy for article quality implies accepting biases that have been amply described and analyzed by bibliometricians (Moed 2002; Van Leeuwen, and Moed 2002; Weingart 2004). However, in the judgment of the authors such biases from the proxy do not significantly alter the methodological logic of this study (as further defined below) or the conclusions that can thus be reached. Given the domain and methods defined for the investigation, it is reasonable to conclude that any errors or limitations related to the proxy do not favour any particular individuals or organizations over others, and thus the statistical validity of the methods can be accepted.

Thus, on the basis of the national context, the availability of data and the other identified study conditions, the following methodology for the study was developed and applied. The *necessary* condition for defining a research unit of an organization as a center of excellence in a given scientific category was made that, among the researchers adhering to the research unit, there must be at least four persons that achieve a qualitative-quantitative scientific production ranking at the level of “top scientist”[4]. The *sufficient* condition for an organizational unit to then be a center of excellence is that its performance ranking is among the highest three[5] in a specific category, at the national level.

The field of observation of the study consists of all public research organizations located in Italy, 676 in total, of which 88 are universities, 68 are public research laboratories, and 530 are research hospitals. The full dataset was constructed from work done in the ORP (Italian Observatory of Public Research)[6], which was set up by the authors. The ORP database lists and classifies the scientific outputs of all researchers appertaining to any of the 676 public research organizations in Italy, drawing on the Thomson Scientific SCI™ (Cd-Rom version), and then ensuring the clear identification of each article listed there, first with unambiguous author names in specific and consistent forms, and then further to the accurate and consistent identification of their home organizations of affiliation in Italy. For this particular study the analysis is limited to the 2001-2003 triennium and to 6 macro-areas: Biology, Chemistry, Earth and space sciences, Engineering, Mathematics, Physics. For the classification of sub-disciplines, the study adopted all the 109 Thomson Scientific categories that can be identified within the 6 macro-areas of the field of observation (see appendix). The procedure for the analyses of scientists and research units follows the 6 steps described below:

Step 1: Ranking of authors in each macro-area on the basis of Scientific Strength (SS), which is the weighted sum of publications realized by the scientist (the weight being the normalized impact factor of the publishing journal)[7] in a given macro-area.

Step 2: Identification of “top scientists”, or in other words, identification of researchers falling in the first decile for Scientific Strength in each macro-area analyzed. In this phase, the authors decided to proceed by identifying top scientists at the level of macro-area, since the choice of a lower level of analysis (that of scientific category, for example) would have led to exclusions of ample numbers of researchers who have notable levels scientific production, but with the production falling in diverse categories within a macro-area. It should also be noted that in Thompson Scientific system of classification it is not rare to find cases where there is strong or partial overlapping, where researchers have scientific production in “adjacent” categories of the same macro-area.

Step 3: Identification of category of specialization of top scientists: analysis of the sectorial distribution of the production of top scientists permitted individuation of each scientist’s category of specialization, as the one Thomson Scientific category with the major concentration of their articles.

Step 4: Identification of exact affiliation of the authors: this task resulted as particularly onerous, since in the SCI™ data base there is no explicit link between the names in the “authors” listing and the organizations indicated in the “addresses” field for the publication. This rendered necessary the development of complex disambiguation algorithms to identify the respective organizational affiliation of each author. The identity of the authors themselves was established as a first step, and a challenging one. In many cases it was actually necessary to manually proceed, case by case, with the recognition of the exact identification of the name and affiliation of the authors, who are, in $SCI^{TM}$ format, indicated only with surname and initial of first name.

Step 5: Clustering of research units: for each category, there followed the identification of all the research units with at least 4 top scientists appertaining to the same organization. These research units are referred to as "top scientist clusters" (TSCs), and are submitted to the subsequent phase of ranking. An analytical check of sensitivity was conducted to assist in selecting and validating this threshold, and revealed that the choice of a higher threshold would have identified research groups affiliated almost exclusively with large organizations.

Step 6: Ranking of identified top scientist clusters on the basis of Fractional Scientific Strength (FSS) of the overall scientific production of the cluster. FSS is analogous to SS but considers the contribution by each scientist to each publication (where contribution, for each publication and for each scientist, is the inverse of the number of co-authors). This indicator, with elimination of double counting due to publications realized in co-authorship by researchers from the same TSC, measures the true scientific strength of the publication portfolio of each TSC.

Step 7: Identification of centers of excellence, as the highest three TSC in terms of FSS, for each scientific category.

The authors wish to note that the 3-year time span of the investigation may be considered too short to assess excellence. Also, as noted above, the authors emphasize that they do not ignore the importance of assumptions concerning the use of the impact factor of a journal as a proxy of the impact/quality of articles published in it. However, the aim of the study is to propose a specific methodology and explore aspects of its implementation, effectiveness and ramifications, with an expectation that these explorations will indicate and stimulate further directions in research.

## RESULTS

### *TOP SCIENTIST CLUSTERS*

For the triennium under observation, there are 51,883 publications represented in the dataset used, with almost 100,000 (Italian) authors. The distribution of publications by macro-area is indicated in Table 1. In terms of output, the area with the greatest number of publications is Physics, with almost 18,000, followed by Biology, at 13,532. Earth and space sciences and Mathematics close up the ranking, with 4,328 and 4,354 publications respectively. In terms of authors, the area with the greatest number is Biology (28,753), followed by Physics (22,641) and Engineering (17,866). The Mathematics area brings up the rear with less than 5% of the authors listed.

As noted above, "top scientists" in any macro-area are identified as those achieving a ranking in the first decile of Scientific Strength out of all authors with publications in that macro-area. As noted from column 5 of Table 1, it was not possible to identify a top scientist cluster (i.e. a group of 4 top scientists falling in the same scientific category affiliated with the same organization) in every scientific category of every macro-area.

In the Biology there were 174 TSC, and they were identified as falling in 10 out of the 28 categories in the scientific macro-area. Engineering registered the highest number of TSC (207), with at least one cluster in a total of 17 of the 39 categories. In Chemistry there was at least one TSC identified in every one of the 9 categories, for a total of 176 clusters. In Physics (191 TSC), there were 5 out of 16 categories with no top scientist clusters. In Earth and space sciences there were 87 TSC identified, in 8 categories out of the total 11 categories. In Mathematics there were a total of 52 TSC identified, found in 5 categories out of the total 6.

***TABLE 1: Distribution of TSC by scientific macro-area***

| Scientific macro-area | N .of categories | N .of publications (2001-2003) | N .of authors | Categories with at least 1 TSC | N .of TSC |
|---|---|---|---|---|---|
| Biology | 28 | 13,532 (21.3%) | 28,753 (29.1%) | 10 | 174 (19.8%) |
| Chemistry | 9 | 12,175 (19.2%) | 16,270 (16.5%) | 9 | 176 (20.1%) |
| Earth and space sciences | 11 | 4328 (6.8%) | 8485 (8.6%) | 8 | 87 (9.9%) |
| Engineering | 39 | 11,216 (17.7%) | 17,866 (18.1%) | 17 | 207 (23.6%) |
| Mathematics | 6 | 4354 (6.9%) | 4769 (4.8%) | 5 | 52 (5.9%) |
| Physics | 16 | 17,873 (28.2%) | 22,641 (22.9%) | 11 | 191 (20.6%) |
| *Total* | *109* | 63,478* | 98,784 | *60* | *877 (100%)* |

** Data include multiple counts due to publications in multi-disciplinary journals*

Looking further into “Earth and space sciences” as an example of a specific macro-area, Table 2 presents the scientific categories identified as including at least one TSC. Almost 80% of the TSC fall in 3 categories: Environmental sciences (21 TSC); Geochemistry and geophysics (21); and “Geosciences, interdisciplinary” (25). In Meteorology and atmospheric sciences and in Mineralogy there were 8 TSC identified, also 2 in Oceanography, but only one in Limnology and water resources. In 3 categories (Geography, physical; Geology; Paleontology) of this macro-area there were no TSC identified. The 87 TSC identified were located in 34 distinct research organizations. The Italian Research Council (CNR), the largest research institution in Italy, registers the greatest number of TSC (19) followed by the National Institute of Geophysics and Vulcanology (INGV), with 7. Figure 1 presents the distribution of top scientist clusters by scientific category. At the head of the ranking, with the most TSC, are the Biochemistry and molecular biology category and the Materials science category, with 64 TSC each. By looking further down the list it can be observed that the first 10 scientific categories in the figure contain half of the all the TSC identified. The number and distribution of clusters are clearly correlated to the distribution of researchers among the categories, or more accurately to the quantity of publications listed for each category[8]. In fact, several observations show this phenomenon: in spite of the Biochemistry & molecular biology category registering a number of publications that is more than double that of Materials science (5,027 versus 2,193), the two categories register exactly the same number of TSC. At the same time, although Spectroscopy presents a mid-range rank of 24$^{th}$ for output among the 60 categories (for a total output of 1,232 articles), it features only 2 TSC. Oceanography also features 2 TSC, but with a production of only 215 publications. Table 3 presents the number of TSC identified per single research organization. The leader is the Italian Research Council, with 100 clusters of top scientists, followed by the Italian Institute for the Physics of Matter, with 99, and the Italian Institute of Nuclear Physics, with 61. First among the universities is Bologna, with 30 TSC, followed by Rome "La Sapienza" (29 TSC) and the University of Milan (23). The first 10 organizations comprise 50% of the total of TSC identified. Bringing up the rear are 38 organizations which together total 56 TSC. The distribution of TSC by typology of their "home" institution (Figure 2) indicates that those within universities compose 59% of the overall total, while those within research hospitals are only 3% of the total.

***TABLE 2: Distribution of TSC per institution in categories of the Earth and space sciences macro-area***

| Research organizations* | Environmental Sciences | Geochemistry and geophysics | Geosciences, interdisciplinary | Limnology | Meteorology and atmospheric sciences | Mineralogy | Oceanography | Water resources | Total |
|---|---|---|---|---|---|---|---|---|---|
| Italian Research Council (CNR) – Bari Research Area | X | - | - | - | - | - | - | - | 1 |
| CNR – Bologna Research Area | X | - | X | - | X | - | - | - | 3 |
| CNR – Florence Research Area | - | X | - | - | - | - | - | - | 1 |
| CNR - La Spezia Research Area | - | - | X | - | - | - | - | - | 1 |
| CNR – Monterotondo Research Area | X | - | - | - | - | - | - | - | 1 |
| CNR – Naples Research Area | X | - | - | - | - | - | - | - | 1 |
| CNR – Palermo Research Area | X | - | - | - | - | - | - | - | 1 |
| CNR – Pallanza Verbania Research Area | X | - | - | X | - | - | - | - | 2 |
| CNR – Pavia Research Area | - | X | - | - | - | X | - | - | 2 |
| CNR – Pisa Research Area | - | X | X | - | - | - | - | - | 2 |
| CNR – Rome Research Area | - | - | X | - | X | - | - | - | 2 |
| CNR – Turin Research Area | - | - | - | - | X | - | - | - | 1 |
| CNR - Venice Research Area | X | - | - | - | - | - | - | - | 1 |
| EU Joint Research Centre ISPRA at Varese | X | - | X | - | X | - | X | - | 4 |
| National Institute of Geophysics and Vulcanology (INGV) – Bologna Research Area | - | - | - | - | X | - | - | - | 1 |
| INGV – Catania Research Area | - | - | X | - | - | - | - | - | 1 |
| INGV – Naples Research Area | - | X | X | - | - | - | - | - | 2 |
| INGV – Palermo Research Area | - | - | X | - | - | - | - | - | 1 |
| INGV – Rome Research Area | - | X | X | - | - | - | - | - | 2 |
| International Center for Theoretical Physics "A. Salam" | - | - | - | - | X | - | - | - | 1 |
| National Institute of Oceanography and Exp. Geophysics | - | X | - | - | - | - | - | - | 1 |
| Polytechnic of Milan | - | X | - | - | - | - | - | - | 1 |
| Polytechnic of Turin | - | - | X | - | - | - | - | X | 2 |
| University "Federico II" | X | X | X | - | - | X | - | - | 4 |
| University Ca' Foscari at Venice | X | - | - | - | - | - | - | - | 1 |
| University of Aquila | - | - | - | - | X | - | - | - | 1 |
| University of Bari | - | - | X | - | - | - | - | - | 1 |
| University of Bologna | X | X | X | - | - | - | - | - | 3 |
| University of Camerino | - | X | - | - | - | - | - | - | 1 |
| University of Catania | - | - | X | - | - | - | - | - | 1 |
| University of Ferrara | - | X | - | - | - | - | - | - | 1 |
| University of Florence | X | X | X | - | - | X | - | - | 4 |
| University of Genoa | X | X | X | - | - | - | - | - | 3 |
| University of Insubria at Varese | X | - | - | - | - | - | - | - | 1 |
| University of Milan | X | X | X | - | - | - | - | - | 3 |
| University of Milan Bicocca | X | - | - | - | - | - | - | - | 1 |
| University of Modena and Reggio Emilia | - | - | - | - | - | X | - | - | 1 |
| University of Padua | - | X | X | - | - | - | - | - | 2 |
| University of Palermo | - | - | X | - | - | - | - | - | 1 |
| University of Pavia | - | - | - | - | - | X | - | - | 1 |
| University of Perugia | - | X | - | - | - | - | - | - | 1 |
| University of Pisa | X | X | X | - | - | X | - | - | 4 |
| University of Rome "La Sapienza" | X | X | X | - | - | - | - | - | 3 |
| University of Rome "Tre" | - | X | X | - | - | - | - | - | 2 |
| University of Siena | X | X | - | - | - | X | - | - | 3 |
| University of Trieste | - | X | X | - | - | - | - | - | 2 |
| University of Turin | X | - | - | - | - | X | - | - | 2 |
| University of Tuscia | X | - | X | - | X | - | - | - | 3 |
| University of Urbino | - | - | X | - | - | - | - | - | 1 |
| University-Polytechnic of the Marche | - | - | - | - | - | - | X | - | 1 |
| *Total* | *21* | *21* | *25* | *1* | *8* | *8* | *2* | *1* | *87* |

** The table gives the names of research organizations in which the TSC are identified: this does not necessarily imply that excellence is identified for the entire research organization.*

***FIGURE 1: Number of TSC per scientific category (in brackets: number of articles per category); data for 2001-2003***

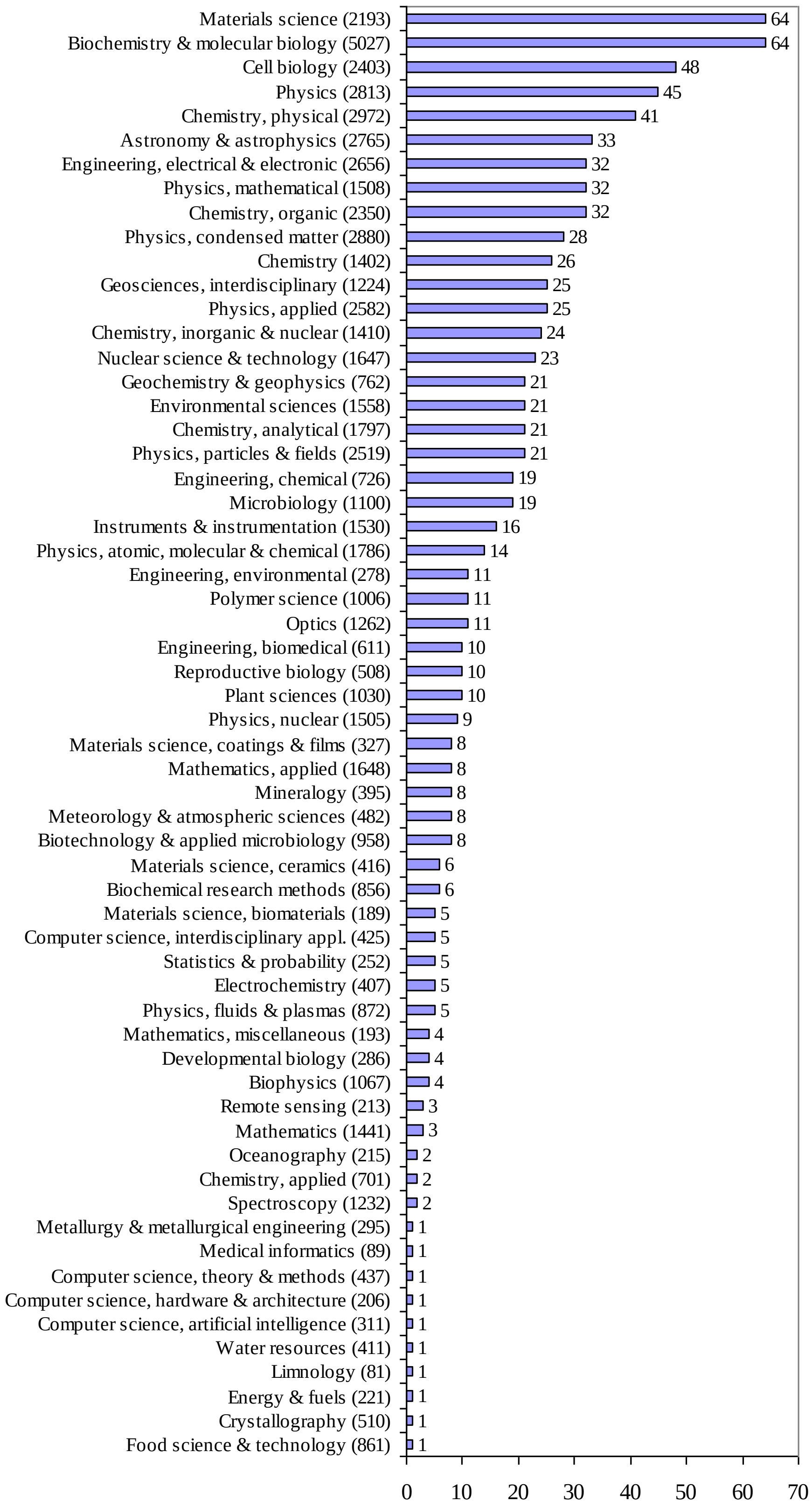

***TABLE 3: Ranking of organizations on the basis of number of identified TSC***

| Research organizations* | Number of TSC | % | Cumulative* (%) |
|---|---|---|---|
| Italian Research Council (CNR) | 110 | 12.5 | 12.5 |
| Italian Institute for the Physics of Matter (INFM) | 99 | 11.3 | 23.8 |
| Italian Institute for Nuclear Physics (INFN) | 61 | 7.0 | 30.8 |
| University of Bologna | 30 | 3.4 | 34.2 |
| University of Rome “La Sapienza” | 29 | 3.3 | 37.5 |
| University of Milan | 23 | 2.6 | 40.1 |
| University of Pisa | 23 | 2.6 | 42.8 |
| University of Florence | 22 | 2.5 | 45.3 |
| University “Federico II” | 19 | 2.2 | 47.4 |
| University of Turin | 19 | 2.2 | 49.6 |
| University of Padua | 18 | 2.1 | 51.7 |
| University of Perugia | 17 | 1.9 | 53.6 |
| Milan Polytechnic | 15 | 1.7 | 55.3 |
| University of Genoa | 15 | 1.7 | 57.0 |
| University of Rome “Tor Vergata” | 15 | 1.7 | 58.7 |
| University of Catania | 14 | 1.6 | 60.3 |
| Turin Polytechnic | 13 | 1.5 | 61.8 |
| University of Padua | 13 | 1.5 | 63.3 |
| University of Trieste | 13 | 1.5 | 64.8 |
| Italian Agency for New Technologies, Energy and the Environment (ENEA) | 12 | 1.4 | 66.1 |
| University of Modena and Reggio Emilia | 12 | 1.4 | 67.5 |
| University of Siena | 12 | 1.4 | 68.9 |
| University of Ferrara | 12 | 1.4 | 70.2 |
| University of Palermo | 11 | 1.3 | 71.5 |
| University of Parma | 11 | 1.3 | 72.7 |
| University of Aquila | 10 | 1.1 | 73.9 |
| University of Messina | 10 | 1.1 | 75.0 |
| University of Rome “Tre” | 10 | 1.1 | 76.2 |
| University of Calabria | 10 | 1.1 | 77.3 |
| National Health Institute (ISS) | 10 | 1.1 | 78.4 |
| University of Bari | 9 | 1.0 | 79.5 |
| University of Salerno | 9 | 1.0 | 80.5 |
| EU Joint Research Centre ISPRA at Varese | 9 | 1.0 | 81.5 |
| National Institute of Geophysics and Vulcanology (INGV) | 7 | 0.8 | 82.3 |
| University of Cagliari | 7 | 0.8 | 83.1 |
| University Ca' Foscari at Venice | 7 | 0.8 | 83.9 |
| University of Milan Bicocca | 7 | 0.8 | 84.7 |
| International School for Advanced Studies, Trieste | 6 | 0.7 | 85.4 |
| University of Verona | 6 | 0.7 | 86.1 |
| University of Camerino | 6 | 0.7 | 86.8 |
| University of Tuscia | 5 | 0.6 | 87.3 |
| University of Sassari | 5 | 0.6 | 87.9 |
| University of Udine | 5 | 0.6 | 88.5 |
| National Electrotechnics Institute “Galileo Ferraris” | 5 | 0.6 | 89.1 |
| Catholic University of the Sacred Heart | 4 | 0.5 | 89.5 |
| University of Eastern Piedmont | 4 | 0.5 | 90.0 |
| University of Brescia | 4 | 0.5 | 90.4 |
| University of Lecce | 4 | 0.5 | 90.9 |
| University of Trento | 4 | 0.5 | 91.3 |
| University of Urbino “Carlo Bo” | 4 | 0.5 | 91.8 |
| University of Insubria at Varese | 4 | 0.5 | 92.2 |
| University-Polytechnic of the Marche | 4 | 0.5 | 92.7 |
| San Raffaele Hospital | 4 | 0.5 | 93.2 |
| International Center for Theoretical Physics “Abdus Salam” | 4 | 0.5 | 93.6 |
| *Others (38)* | *56* | *6.4* | *100.0* |
| *Total* | *877* | | |

** The table gives the names of research organizations in which the TSC are identified: this does not necessarily imply that excellence is identified for the entire research organization*

***FIGURE 2: Distribution of TSC by typology of institution***

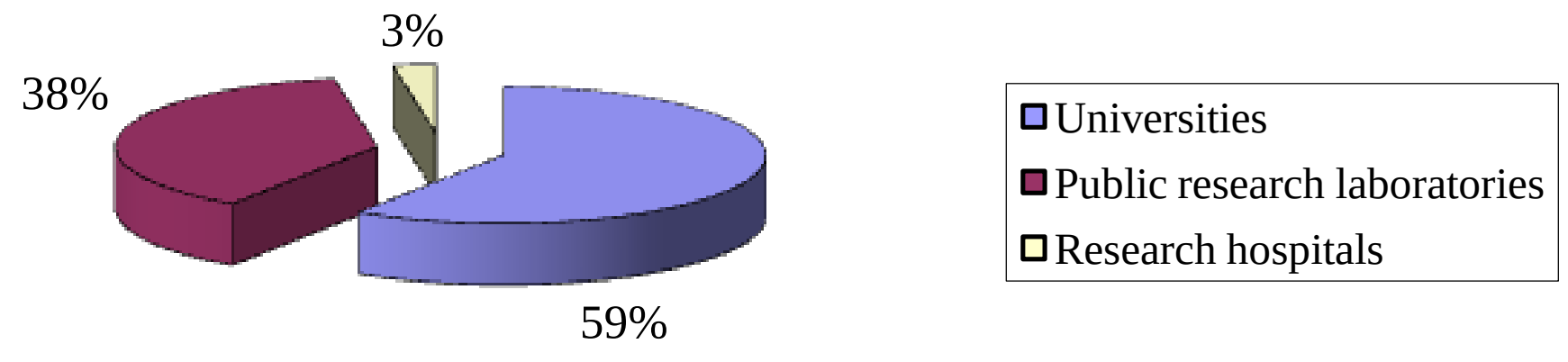

## *CENTERS OF EXCELLENCE*

Centers of excellence (COE) are defined as the first three top scientist clusters found in each scientific category, when ranked by Fractional Scientific Strength. Figure 3 shows their distribution through the 6 scientific macro-areas under observation. In total, there were 157 COE[9] identified, of which 51.6% appertain to only 8 research organizations. The distribution by typology of institution is presented in Figure 4. 63% of the identified centers of excellence are situated in universities, 34% in public research laboratories and 3% in research hospitals. Table 4 ranks the organizations by number of centers of excellence. There are 42 organizations with at least one COE. The institution with the most COEs is again the Italian Research Council, with 19 centers, followed by the University of Bologna (12 COE). University of Rome "La Sapienza" has 11 centers of excellence, the same number as registered for the Italian Institute for the Physics of Matter. The major part of the centers of excellence are situated in organizations of substantial size, but smaller organizations with concentrations of CEO are not infrequent. For example there are 5 centers of excellence at the comparatively small EU Joint Research Centre ISPRA.

***FIGURE 3: Distribution of centers of excellence by macro-area***

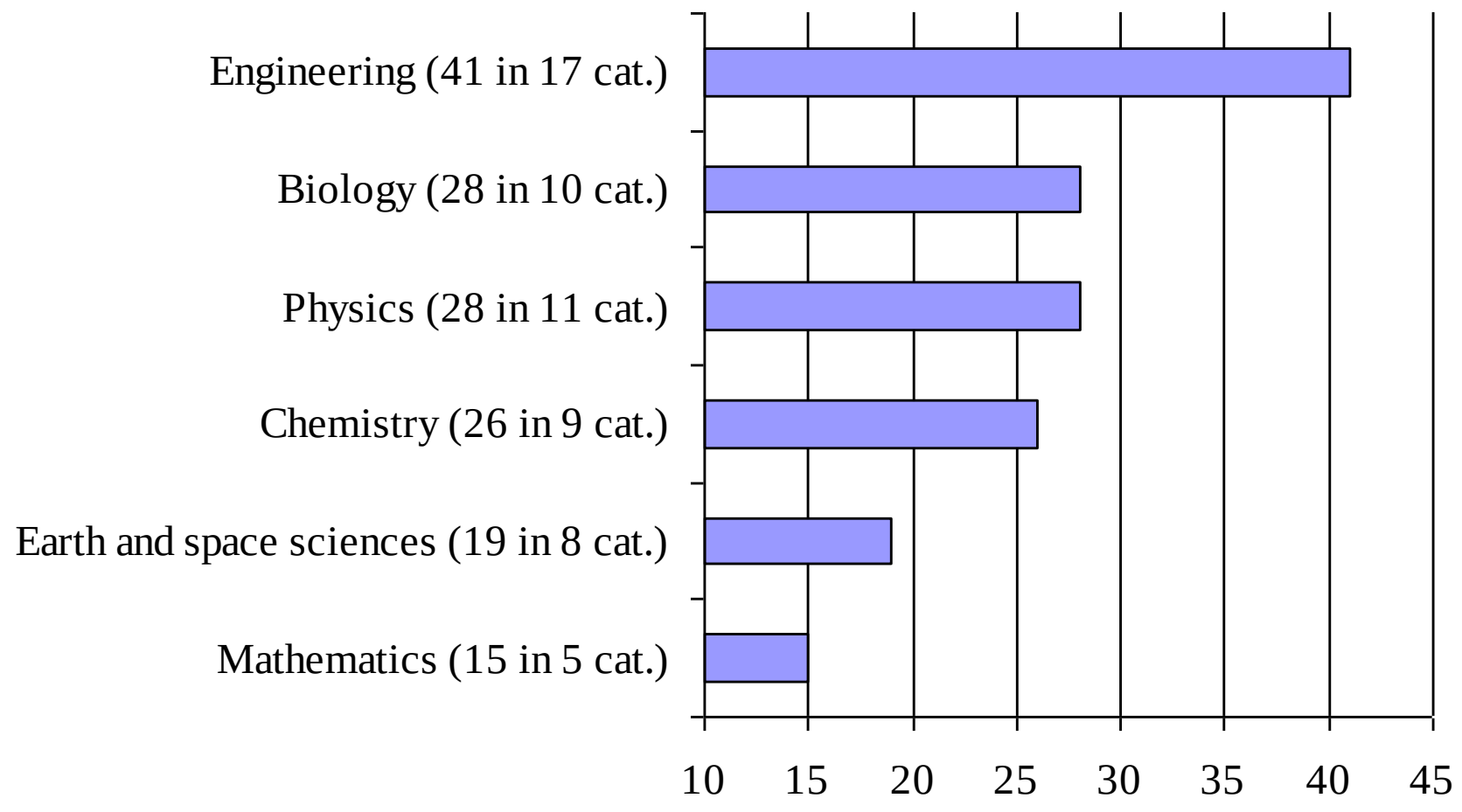

***FIGURE 4: Distribution of centers of excellence by typology of institution***

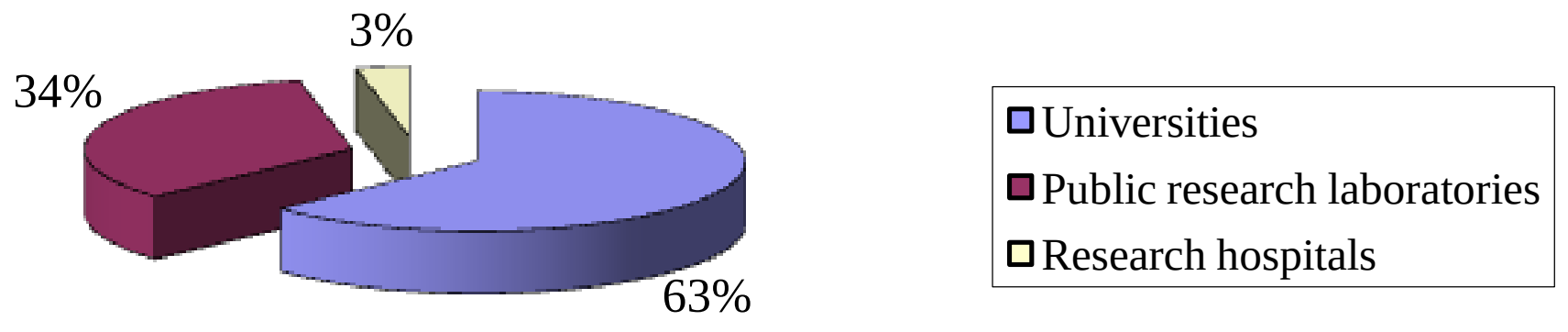

***TABLE 4: Ranking of organizations by number of centers of excellence***

| Research organizations* | Number of COE | % | Cumulative* (%) |
|---|---|---|---|
| Italian Research Council (CNR) | 19 | 12.1 | 12.1 |
| University of Bologna | 12 | 7.6 | 19.7 |
| University of Rome “La Sapienza” | 11 | 7.0 | 26.8 |
| Italian Institute for the Physics of Matter | 11 | 7.0 | 33.8 |
| Polytechnic of Milan | 8 | 5.1 | 38.9 |
| University of Pisa | 8 | 5.1 | 43.9 |
| University “Federico II” | 6 | 3.8 | 47.8 |
| Italian Institute for Nuclear Physics | 6 | 3.8 | 51.6 |
| EU Joint Research Centre ISPRA at Varese | 5 | 3.2 | 54.8 |
| Polytechnic of Turin | 5 | 3.2 | 58.0 |
| University of Milan | 5 | 3.2 | 61.1 |
| University of Padua | 5 | 3.2 | 64.3 |
| University of Florence | 4 | 2.5 | 66.9 |
| University of Rome “Tor Vergata” | 4 | 2.5 | 69.4 |
| University of Turin | 4 | 2.5 | 72.0 |
| Italian Agency for New Technologies, Energy and the Environment | 4 | 2.5 | 74.5 |
| University of Trieste | 3 | 1.9 | 76.4 |
| University of Salerno | 3 | 1.9 | 78.3 |
| National Institute of Geophysics and Volcanology | 3 | 1.9 | 80.3 |
| Rizzoli Orthopaedic Institute, Bologna | 2 | 1.3 | 81.5 |
| International Center for Theoretical Physics “Abdus Salam” | 2 | 1.3 | 82.8 |
| San Raffaele Hospital | 2 | 1.3 | 84.1 |
| University of Bari | 2 | 1.3 | 85.4 |
| University of Genoa | 2 | 1.3 | 86.6 |
| University of Verona | 2 | 1.3 | 87.9 |
| University of Catania | 2 | 1.3 | 89.2 |
| University-Polytechnic of the Marche | 2 | 1.3 | 90.4 |
| National Health Institute | 1 | 0.6 | 91.1 |
| National Cancer Institute | 1 | 0.6 | 91.7 |
| Astrophysical Observatory at Arceteri | 1 | 0.6 | 92.4 |
| Astronomical Observatory at Brera | 1 | 0.6 | 93.0 |
| University Ca' Foscari at Venezia | 1 | 0.6 | 93.6 |
| University of Tuscia | 1 | 0.6 | 94.3 |
| University of Messina | 1 | 0.6 | 94.9 |
| University of Modena and Reggio Emilia | 1 | 0.6 | 95.5 |
| University of Parma | 1 | 0.6 | 96.2 |
| University of Padua | 1 | 0.6 | 96.8 |
| University of Perugia | 1 | 0.6 | 97.5 |
| University of Siena | 1 | 0.6 | 98.1 |
| University of Trento | 1 | 0.6 | 98.7 |
| University of Rome “Tre” | 1 | 0.6 | 99.4 |
| University of Ferrara | 1 | 0.6 | 100.0 |
| *Total* | *157* | | |

** The table gives the names of research organizations in which the top scientist clusters are identified: this does not necessarily imply that excellence is identified for the entire research organization*

Table 5 and Table 6 illustrate the identification of centers of excellence from the larger listings of top scientist clusters. Table 5 describes the case of the National Institute of Geophysics and Volcanology, in which there are 7 TSC, of which 3 are classified as centers of excellence, all in the macro-area of Earth and space sciences. Table 6 ranks all Italian organizations identified as having clusters of top scientists in the area “Engineering, biomedical”. At the top of the ranking are three organizations that emerge as centers of excellence: the Rizzoli Orthopaedic Institute (Bologna), with a Fractional Scientific Strength of 75.97, Milan Polytechnic (FSS 27.53) and the University of Bologna (22.70).

***TABLE 5: COE and TSC situated within the "National Institute of Geophysics and Volcanology"***

| Category | Research Unit | COE/TSC | Top scientists |
|---|---|---|---|
| Geochemistry & geophysics | Naples | COE | Chiodini G; Ventura G; Caliro S; Russo M |
| " | Rome | COE | Boschi E; Sagnotti L; Florindo F; Speranza F |
| Geosciences, interdisciplinary | Rome | COE | Piersanti A; Cocco M; Marra F; Chiarabba C |
| " | Naples | TSC | Delpezzo E; Saccorotti G; Denatale G; Macedonio G |
| " | Palermo | TSC | Italiano F; Favara R; Sortino F; Caracausi A |
| " | Catania | TSC | Puglisi G; Neri M; Andronico D; Oppenheimer C |
| Meteorology & atmospheric sciences | Bologna | TSC | Navarra A; Delecluse P; Gualdi S; Guilyardi E |

***TABLE 6: COE and TSC for the category "Engineering, biomedical"***

| Research organizations* | COE/TSC | FSS | Top scientists |
|---|---|---|---|
| Rizzoli Orthopaedic Institute, Bologna | COE | 75.97 | Ciapetti G; Granchi D; Cenni E; Toni A |
| Polytechnic of Milan | COE | 27.53 | Pedotti A; Pietrabissa R; Fumero R; Baselli G |
| University of Bologna | COE | 22.70 | Facchini A; Taddei P; Cappello A; Biffi M |
| National Health Institute | TSC | 21.38 | Barbaro V; Daniele C; Davenio G; Grigioni M |
| University of Padua | TSC | 21.14 | Cobelli C; Sparacino G; Abatangelo G; Brun P |
| University of Turin | TSC | 16.24 | Costa L; Bracco P; Masse A; Jacobson K |
| Polytechnic of Turin | TSC | 15.55 | Verne E; Farina D; Brovarone CV; Merletti R |
| University of Milan | TSC | 11.21 | Porta A; Malliani A; Pagani M; Chiapasco M |
| University of Perugia | TSC | 7.67 | Becchetti E; Belcastro S; Lilli C; Locci P |
| Italian Research Council | TSC | 5.43 | Clemente F; Ferrari G; Guaragno M; Delazzari C |

** The table gives the names of research organizations in which the TSC are identified: this does not necessarily imply that excellence is identified for the entire research organization*

## GEOGRAPHIC DISTRIBUTION OF CENTERS OF EXCELLENCE

Analysis of the geographic distribution of centers of excellence can offer information relevant to both national and regional policy makers. For example, in Italy, where 50.7% of R&D spending is financed by government (OECD 2007), mapping of centers of scientific excellence is undoubtedly useful in the process of formulating and monitoring policies for territorial re-balancing and regional development.

Table 7, for each scientific macro-area, indicates the number and the percentage incidence of centers of excellence as situated in each Italian geographic macro-area. The distribution is relatively uniform, with prevalence towards locations in the central administrative regions, where there are 52 centers of excellence, representing a third of the national total. The most prominent characteristic of the data is the concentration of centers of excellence in the region of Lazio, which registers 29 COE (Table 8). This result confirms expectations, Lazio being the region with the highest concentration of public spending on R&D (25.4% of the national total) and of "public" researchers (27.0% of the national total, including 13.8% of university researchers and 51.9% of those in public research laboratories), ISTAT (2005). Half of the national centers of excellence are situated in northern Italy, with a slight prevalence for the north western area (44 COE) with respect to the north-east (38). In the south there are 23 centers of excellence, amounting to 14.65% of the total. Analyses of individual scientific macro-

areas reveal some differentiations with respect to the data for the overall distribution of all COE. For example, of the 26 Chemistry centers of excellence in the nation, 6 (23%) are situated in southern administrative regions and only 3 are situated in the north-west. Centers of excellence for the Physics macro-area are concentrated in a significant manner in the central regions. In southern Italy there were no centers of excellence identified for Mathematics.

***TABLE 7: Distribution of COE by scientific macro-area and geographic macro-area***

| Scientific macro-area | North-west | North-east | Center | South | Total |
|---|---|---|---|---|---|
| Biology | 8 (28.6%) | 6 (21.4%) | 9 (32.1%) | 5 (17.9%) | 28 (17.8%) |
| Chemistry | 3 (11.5%) | 9 (34.6%) | 8 (30.8%) | 6 (23.1%) | 26 (16.6%) |
| Earth and space sciences | 7 (36.8%) | 4 (21.1%) | 6 (31.6%) | 2 (10.5%) | 19 (12.1%) |
| Engineering | 14 (33.3%) | 10 (23.8%) | 13 (31.0%) | 5 (11.9%) | 42 (26.8%) |
| Mathematics | 5 (33.3%) | 5 (33.3%) | 5 (33.3%) | 0 (0.0%) | 15 (9.6%) |
| Physics | 7 (25.9%) | 4 (14.8%) | 11 (40.7%) | 5 (18.5%) | 27 (17.2%) |
| *Total* | *44 (28.0%)* | *38 (24.2%)* | *52 (33.1%)* | *23 (14.6%)* | *157 (100%)* |

Table 8 presents data on a regional basis. Lazio and Lombardy lead all of the 20 national regions with 127 and 125 top scientist clusters respectively, situated in their territories. The COE are instead equal in number, at 29 for both of these regions. Following up are Emilia Romagna and Tuscany, with 105 and 94 TSC respectively, and sharing the same number of centers of excellence (19). With the exception of Valle d'Aosta (where the first university has only recently been founded and where there are no other public research laboratories), the research methodologies permitted the identification of at least one TSC in each of the nation's regions and at least one COE in 15 regions. In this regard, Abruzzo, Calabria and Sardinia, though with a fair number of TSC (13, 13 and 14), do not have any centers of excellence. Another two regions of the south, Apulia and Sicily, also with a good number of TSC (30 and 54 respectively), register very few centers of excellence (4 and 5).

In general, a picture emerges of a distribution of centers of excellence that retraces the regional distribution of researchers: the index of correlation between data for the regional distribution of "public" researchers and regional distribution of centers of excellence is, in fact, equal to 0.82.

Table 9 presents the distribution of centers of excellence in Italy's regions, in terms of the percentage incidence for each scientific macro-area. Glancing down the column rows, Lombardy possesses the highest percentage (25%) of national COE in Biology, and the same region is also first for 3 other macro-areas: Earth and space sciences (together with Lazio), Engineering, Mathematics. In Chemistry the top ranking is shared by Emilia Romagna, Lazio and Campania. In Physics, Tuscany has the highest presence of centers of excellence.

Table 10 gives a differing presentation of the distribution, now examining the totals of COE identified in each region and identifying percentages attributable to each scientific macro-area. A glance across the table rows shows the macro-area strong points of each region.

***TABLE 8: Distribution of TSC and COE by region***

| Region | Geographic area | TSC | COE |
|---|---|---|---|
| Piedmont | North-west | 59 (6.7%) | 12 (7.6%) |
| Valle d'Aosta | North-west | 0 | 0 |
| Lombardy | North-west | 127 (14.5%) | 29 (18.5%) |
| Liguria | North-west | 30 (3.4%) | 3 (1.9%) |
| Trentino Alto Adige | North-east | 11 (1.3%) | 1 (0.6%) |
| Veneto | North-east | 56 (6.4%) | 13 (8.3%) |
| Friuli Venezia Giulia | North-east | 44 (5.0%) | 5 (3.2%) |
| Emilia Romagna | North-east | 105 (12.0%) | 19 (12.1%) |
| Tuscany | Center | 94 (10.7%) | 19 (12.1%) |
| Umbria | Center | 18 (2.1%) | 1 (0.6%) |
| Marche | Center | 21 (2.4%) | 3 (1.9%) |
| Lazio | Center | 124 (14.1%) | 29 (18.5%) |
| Abruzzo | Center | 13 (1.5%) | 0 (0.0%) |
| Molise | South | 1 (0.1%) | 0 (0.0%) |
| Campania | South | 60 (6.8%) | 13 (8.3%) |
| Apulia | South | 30 (3.4%) | 4 (2.5%) |
| Basilicata | South | 3 (0.3%) | 1 (0.6%) |
| Calabria | South | 13 (1.5%) | 0 (0.0%) |
| Sicily | South | 54 (6.2%) | 5 (3.2%) |
| Sardinia | South | 14 (1.6%) | 0 (0.0%) |
| *Total* | | *877 (100.0%)* | *157 (100.0%)* |

***TABLE 9: Distribution of COE across the regions, as percentage of scientific macro-area totals***

| Region | Biology | Chemistry | Earth and space sciences | Engineering | Mathematics | Physics |
|---|---|---|---|---|---|---|
| Piedmont | 3.6 | 3.8 | 15.8 | 9.5 | 6.7 | 7.4 |
| Valle d'Aosta | 0.0 | 0.0 | 0.0 | 0.0 | 0.0 | 0.0 |
| Lombardy | 25.0 | 3.8 | 21.1 | 21.4 | 26.7 | 14.8 |
| Trentino Alto Adige | 0.0 | 0.0 | 0.0 | 2.4 | 0.0 | 0.0 |
| Veneto | 7.1 | 7.7 | 5.3 | 7.1 | 13.3 | 11.1 |
| Friuli Venezia Giulia | 0.0 | 7.7 | 10.5 | 0.0 | 0.0 | 3.7 |
| Liguria | 0.0 | 3.8 | 0.0 | 2.4 | 0.0 | 3.7 |
| Emilia Romagna | 14.3 | 19.2 | 5.3 | 14.3 | 20.0 | 0.0 |
| Tuscany | 10.7 | 11.5 | 5.3 | 9.5 | 13.3 | 22.2 |
| Umbria | 0.0 | 0.0 | 0.0 | 2.4 | 0.0 | 0.0 |
| Marche | 3.6 | 0.0 | 5.3 | 0.0 | 0.0 | 3.7 |
| Lazio | 17.9 | 19.2 | 21.1 | 19.0 | 20.0 | 14.8 |
| Abruzzo | 0.0 | 0.0 | 0.0 | 0.0 | 0.0 | 0.0 |
| Molise | 0.0 | 0.0 | 0.0 | 0.0 | 0.0 | 0.0 |
| Campania | 10.7 | 19.2 | 10.5 | 4.8 | 0.0 | 3.7 |
| Apulia | 3.6 | 0.0 | 0.0 | 2.4 | 0.0 | 7.4 |
| Basilicata | 0.0 | 0.0 | 0.0 | 2.4 | 0.0 | 0.0 |
| Calabria | 0.0 | 0.0 | 0.0 | 0.0 | 0.0 | 0.0 |
| Sicily | 3.6 | 3.8 | 0.0 | 2.4 | 0.0 | 7.4 |
| Sardinia | 0.0 | 0.0 | 0.0 | 0.0 | 0.0 | 0.0 |
| *Total* | *100.0* | *100.0* | *100.0* | *100.0* | *100.0* | *100.0* |

*TABLE 10: Distribution of COE by macro-area, as percentage of the total centers of excellence in each region*

| Region | Biology | Chemistry | Earth and space sciences | Engineering | Mathematics | Physics | Total |
|---|---|---|---|---|---|---|---|
| Piedmont | 8.3 | 8.3 | 25.0 | 33.3 | 8.3 | 16.7 | 100 |
| Valle d'Aosta | - | - | - | - | - | - | - |
| Lombardy | 24.1 | 3.4 | 13.8 | 31.0 | 13.8 | 13.8 | 100 |
| Trentino Alto Adige | 0.0 | 0.0 | 0.0 | 100.0 | 0.0 | 0.0 | 100 |
| Veneto | 15.4 | 15.4 | 7.7 | 23.1 | 15.4 | 23.1 | 100 |
| Friuli Venezia Giulia | 0.0 | 40.0 | 40.0 | 0.0 | 0.0 | 20.0 | 100 |
| Liguria | 0.0 | 33.3 | 0.0 | 33.3 | 0.0 | 33.3 | 100 |
| Emilia Romagna | 21.1 | 26.3 | 5.3 | 31.6 | 15.8 | 0.0 | 100 |
| Tuscany | 15.8 | 15.8 | 5.3 | 21.1 | 10.5 | 31.6 | 100 |
| Umbria | 0.0 | 0.0 | 0.0 | 100.0 | 0.0 | 0.0 | 100 |
| Marche | 33.3 | 0.0 | 33.3 | 0.0 | 0.0 | 33.3 | 100 |
| Lazio | 17.2 | 17.2 | 13.8 | 27.6 | 10.3 | 13.8 | 100 |
| Abruzzo | - | - | - | - | - | - | - |
| Molise | - | - | - | - | - | - | - |
| Campania | 23.1 | 38.5 | 15.4 | 15.4 | 0.0 | 7.7 | 100 |
| Apulia | 25.0 | 0.0 | 0.0 | 25.0 | 0.0 | 50.0 | 100 |
| Basilicata | 0.0 | 0.0 | 0.0 | 100.0 | 0.0 | 0.0 | 100 |
| Calabria | - | - | - | - | - | - | - |
| Sicily | 20.0 | 20.0 | 0.0 | 20.0 | 0.0 | 40.0 | 100 |
| Sardinia | - | - | - | - | - | - | - |

# SUMMARY AND CONCLUSIONS

Among scholars, there is a growing impetus towards a shared definition of scientific excellence, along with methods for its identification and evaluation. The commitment of scholars reflects the importance and relevance of the theme, which is of interest even to those not directly involved in the field of research policy, including to the general public. The present work offers a methodological contribution to a debate already well under way. It is a methodology for identifying the centers of excellence of a national research system, which is in keeping with objectives and requisites set in by the European Commission.

The methodology is based on a fundamental definition of "center of excellence", developed to be compatible with the context of the study. A center of excellence for a specific scientific category is definable as an organizational group of staff including a minimum nucleus (4) of top scientists. Top scientists are in turn identified as those falling in the first decile of national rankings of a bibliometric indicator (Scientific Strength) which can be applied to the measurement of scientific production and its potential scientific impact, in every scientific macro-area.

The proposed methodology was tested on the entirety of 6 scientific macro-areas of the Italian public research system (676 organizations) and permitted the tentative identification of 877 clusters of top scientists and 157 centers of excellence. Unlike others in the literature, this particular study is characterized by its exhaustive field of investigation, concerning organizations throughout the nation, and with the wide range and number of scientific disciplines being examined. Again unlike others, this study also begins with and executes a complete census of excellence at the level of individual research groups, rather than only presenting an examination at the aggregate level of the whole organizations.

The authors carried out this exploratory study with certain limitations on resources,

such as data that were limited to a three-year-time period and a lack of availability of citations of articles. However the authors are confident that by expanding the period of observation to between 5 and 7 years and by directly measuring the quality of publications through the frequency of citations, rather than through the impact factor of the publishing journals, the measures of productivity will become more robust and reliable in identifying clear differences between research groups. At that point the findings from the methodology can be accepted as no longer indicative, but immediately useful towards decisions of policy and action.

Though further refinement will arrive, this first exploratory approach already opens the road to new forms of analyses, such as at the levels of institutions, disciplines or geographic and administrative areas. The proposed methodology is clearly adaptable to a variety of contexts, in both national and supranational research systems. It offers assistance to single researchers and administrators, to the private sector, and to national and European Union policy makers.

## NOTES

[1] The *Science Citation Index* (SCI®-Cd Rom version) lists bibliographic information concerning 3,700 of the world's leading scholarly science and technical journals, covering approximately 180 scientific categories grouped in 8 macro-areas.

[2] In this investigation, the only research outputs measured are scientific journal publications. Other forms of outputs (such as proceedings, monographs, patents or prototypes) are excluded from measurement. However in the scientific categories examined here, journal publications are highly representative of real output from all research activity.

[3] For a given year, the impact factor of a journal is calculated by dividing the "number of "current-year" citations to source items published in the journal" by "the number of items published in the journal during the previous two years".

[4] The term "top scientist" is applied, as explained in the following paragraphs, on the basis of the quantity and quality of scientific publications realized in the triennium under observation, in a given scientific macro-area.

[5] This condition is, in part, adopted by the authors for clarity in discussion of the present investigations.: In the ranking of research organizational units, the performance difference between the first three organizational units and the following ones could be minimal. Thus the dividing line between those defined as centers of excellence and those that do not attain the definition could be somewhat arbitrary. However we expect that further research, over an extended period of observation, adding further measurement of numbers of citations, would make the differences in observed results between research units much larger and more distinctive.

[6] The full details of the specific ORP methodologies used to construct the dataset are described in Abramo, D'Angelo, and Pugini 2008.

[7] The distribution of impact factors of journals differs substantially from one category to another. The normalization of each journal's impact factor with respect to the category average permits limiting the distortions in comparing performances between different categories. More about the effects of field normalization may be found in Zitt, Ramanana-Rahari, and Bassecoulard 2005.

[8] The index of correlation between number of publications and number of TSC per sector equals 0.84.

[9] In a number of categories, the number of top scientist clusters identified is less than or equal to 3. In these cases, all the clusters identified were considered centers of excellence.

# APPENDIX

## LIST OF SCIENTIFIC CATEGORIES CONSIDERED IN THE STUDY

| **Macro-area** | **Category** |
|---|---|
| Biology | Agricultural engineering - Agricultural multidisciplinary - Agriculture - Agriculture, dairy and animal science - Agriculture, soil science - Biochemical research methods - Biochemistry and molecular biology - Biodiversity conservation - Biology - Biophysics - Biotechnology and applied microbiology - Cell biology - Developmental biology - Ecology - Entomology - Evolutionary biology - Fisheries - Food science and technology - Forestry - Horticulture - Marine and freshwater biology - Microbiology - Mycology - Ornithology - Plant sciences - Reproductive biology - Veterinary sciences - Zoology |
| Chemistry | Chemistry - Chemistry, analytical - Chemistry, applied - Chemistry, inorganic and nuclear - Chemistry, organic - Chemistry, physical - Electrochemistry - Physics, atomic, molecular and chemical - Polymer science |
| Earth and space sciences | Environmental sciences - Geochemistry and geophysics - Geography, physical - Geology - Geosciences, interdisciplinary - Limnology - Meteorology and atmospheric sciences - Mineralogy - Oceanography - Paleontology - Water resources |
| Engineering | Aerospace engineering and technology - Computer science, artificial intelligence - Computer science, cybernetics - Computer science, hardware and architecture - Computer science, information systems - Computer science, interdisciplinary applications - Computer science, software, graphics, programming - Computer science, theory and methods - Construction and building technology - Engineering - Engineering ocean - Engineering, biomedical - Engineering, chemical - Engineering, civil - Engineering, electrical and electronic - Engineering, environmental - Engineering, geological - Engineering, industrial - Engineering, manufacturing - Engineering, mechanical - Engineering, petroleum - Instruments and instrumentation - Materials science - Materials science, biomaterials - Materials science, ceramics - Materials science, characterization and testing - Materials science, coatings and films - Materials science, composites - Materials science, paper and wood - Materials science, textiles - Medical informatics - Metallurgy and metallurgical engineering - Mining and mineral processing - Nuclear science and technology - Remote sensing - Robotics - Robotics and automatic control - Telecommunications - Transportation science and technology |
| Mathematics | Mathematics - Mathematics, applied - Mathematics, miscellaneous - Operations research and management science - Physics, mathematical - Statistics and probability |
| Physics | Acoustics - Astronomy and astrophysics - Crystallography - Energy and fuels - Mechanics - Microscopy - Optics - Photographic technology - Physics - Physics, applied - Physics, condensed matter - Physics, fluids and plasmas - Physics, nuclear - Physics, particles and fields - Spectroscopy - Thermodynamics |